\documentclass[prb,twocolumn,groupedaddress,showpacs,superscriptaddress]{revtex4}

\usepackage{graphicx,amscd,amsmath,amssymb}

\usepackage{graphicx}
\usepackage{dcolumn}
\usepackage{bm}
\bibstyle{apsrev}

\begin{document}

\title{High field electro-thermal transport in metallic carbon nanotubes}


\author{Marcelo A. Kuroda}
\affiliation{Dept.~of Physics and Beckman Institute, University of
Illinois, Urbana-Champaign, IL 61801} \email{mkuroda@illinois.edu}
\author{Jean-Pierre Leburton}
\affiliation{Dept.~of Electrical and Computer Engineering, Department of Physics and Beckman Institute, University of
Illinois, Urbana-Champaign, IL 61801} \email{jleburto@illinois.edu}
\date{\today}

\begin{abstract}
We describe the electro-thermal transport in metallic carbon nanotubes (m-CNTs) by a semi-classical approach that takes into account the high-field dynamical interdependence between charge carrier and phonon populations. Our model is based on the self-consistent solution of the Boltzmann transport equation and the heat equation mediated by a phonon rate equation that accounts for the onset of non-equilibrium (optical) phonons in the high-field regime. Given the metallic nature of the nanostructures, a key ingredient of the model is the assumption of local thermalization of charge carriers. Our theory remarkably reproduces the room temperature electrical characteristics of m-CNTs on substrate and free standing (suspended), shedding light on charge-heat transport in these one dimensional nanostructures. In particular, the negative differential resistance observed in suspended m-CNTs under electric stress is attributed to inhomogeneous field profile induced by self-heating rather than the presence of hot phonons.
\end{abstract}


\maketitle

\section{Introduction}

Since their discovery \cite{iijima1991}, carbon nanotubes (CNTs) have been subject of intense research for both scientific interest and technological purpose. Their thermal, mechanical and electrical properties \cite{dresselhaus} make them prominent candidates for broad range of applications in nanotechnology \cite{baughman2002}. The electronic structure of these one-dimensional (1D) macromolecules exhibit semiconducting or metallic (m) features depending on their chirality \cite{saito1992}. The latter have been proposed for interconnect applications because they can support currents densities larger than copper without suffering electro-migration \cite{mceuen2002,kreupl2002}. Recent experiments on individual m-CNTs  in the high field regime \cite{yao2000, park2004, javey2004, pop2005, sundqvist2007} have revealed strong nonlinear transport characteristics, such as current saturation and negative differential resistance, which have been attributed to the emergence of optical phonon (OP) populations out of equilibrium (hot phonons)\cite{pop2005,lazzeri2006}. Additionally, the large electrical power dissipated in the high bias regime have fostered further studies of the electro-thermal transport in these structures \cite{maune2006, pop2007, hsu2008, avedisian2008, deshpande2009, shi2009}. As a consequence numbers of theoretical and experimental works have attempted to quantify both the electron-phonon (e-p) and phonon-phonon (p-p) coupling interaction in m-CNTs \cite{lazzeri2005,oron2008,pennington2008,song2008} .

Prominent among the theoretical approaches for investigating high-field transport in m-CNTs at room temperature are the Boltzmann transport equation (BTE) \cite{kuroda2005,auer2006}, Monte Carlo simulations \cite{javey2004}, Landauer-Buttiker formalism \cite{pop2005} based on the Fermi golden's rule \cite{lazzeri2006} that successfully described transport in nanoscale semiconductors. However one of the major drawbacks of these approaches is the absence of electron-electron (e-e) interaction, despite the metallic character of the system, and the contribution of the charge carriers to thermal transport. In this work, which is applicable to any metallic 1D system, we describe the system dynamics as the interplay between carrier, acoustic phonon (AP) and OP populations. Our self-consistent approach describing electro-thermal transport accounts for the onset of phonon and carrier populations out of equilibrium and provides a comprehensive picture of the carrier flow and heat exchange in the nanostructures. Our model is in excellent agreement with experiments on individual m-CNTs in both substrate supported (SS) and free standing (FS) configuration at room temperature.

\section{Model}

We describe the high-field transport in m-CNTs by accounting for both electric and thermal transport through the 1D structure. While the former originates from the charge carrier populations only, both electron and phonon populations contribute to heat transfer in a conjugated manner. 

\subsection{Electric and energy flow}

\begin{figure}
     \centering
     \includegraphics[width=3.25in]{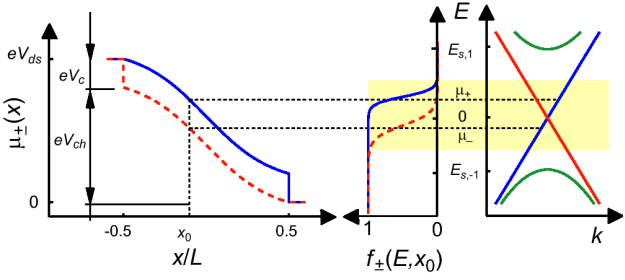}
     \caption{Left: Schematics of the forward and backward quasi-Fermi level profiles as a function of the distance along the m-CNT. Center: Forward and backward distribution functions at position $x=x_0$. Right: Band structure at $x=x_0$. Contributions of upper and lower subbands are neglected as they lie far from the active scattering area (shaded region).  \label{fig:scheme_dist} }
\end{figure}

In small diameter m-CNTs, the higher conduction and lower valence subbands lie far above or below the Fermi level so that the charge flow arises solely from the contribution of the conducting bands crossing the Fermi level (here taken as the zero-energy level), which are well described by a linear $E-k$ dispersion:
\begin{equation}
E_\pm(k)= \pm \hbar v_F k \label{eq:bandst}
\end{equation}
where $v_F$ is the Fermi velocity ($v_F\approx 8\times10^7$cm/sec in CNTs) \cite{saito1992} and the + ($-$) sign corresponds to the forward (backward) carrier branches. Each of these branches has a degeneracy $2 g_c$ (where the factor of 2 indicates the spin degeneracy and $g_c$ = 2 for m-CNTs). Due to the efficient intrabranch e-e interaction \cite{kuroda2008} carrier populations are thermalized and described by Fermi-Dirac distribution functions. In the presence of an external electric field $F$, the local quasi-Fermi levels of forward and backward populations separate ($\mu_+\neq \mu_-$), and the carrier distributions in each branch are not necessarily in local thermal equilibrium with each other ($T_+(x)\neq T_-(x)$), as illustrated in Fig.~\ref{fig:scheme_dist}. The electric current in m-CNTs reads \cite{kuroda2005}:
\begin{equation}
I = g_c G_0 \frac{\left(\mu_+-\mu_-\right)}{e}
\end{equation}
where $e$ is the electron charge and $G_0 = e^2/(\pi \hbar)$ is the quantum conductance. Here we point out that the quasi-Fermi level difference does not necessarily coincide with the drain-source voltage, i.e.~$|\mu_+-\mu_- |\leq |e V_{ds}|$, where the equality only holds in the absence of back-scattering. In other words carriers injected from the contacts are immediately thermalized by the effective e-e interaction. Hence, even if $eV_{ds}$ is larger than the separation between subbands, contribution of conducting subbands can be neglected provided that the local quasi-Fermi levels in each branch lie far away from the upper or lower subband ($|E_s - \mu_\eta|\gg k_BT_\eta$), as illustrated in Fig.~\ref{fig:scheme_dist}. We also recognize that quasi-Fermi levels $\mu_\pm(x)$ are position-dependent, but their difference remains constant for current conservation \cite{kuroda2005}.

In this metallic system, the distribution function $f_\pm(E,x)$ for each branch obeys the stationary BTE expressed as:
\begin{equation}
\pm v_F\left[ \,\partial_x f_\pm(E,x) + eF  \partial_E
f_\pm(E,x) \right]=\partial_t f_\pm\Big\arrowvert_{coll}, \label{eq:boltzmann}
\end{equation}
after using the linear dispersion (Eq.~\ref{eq:bandst}). Here, $x$ denotes the position along the m-CNT and the right hand side of the equation is the collision integral. The local temperatures and quasi-Fermi levels of the thermalized distributions can be determined using the method of moments (from here on we omit the variable $x$ for sake of brevity). We have recently shown that the electric field $F$ and carrier temperature $T_\pm$ profiles fulfill the relations \cite{kuroda2008}:
\begin{equation}
-e F  = \pm
\frac{\pi\hbar}{g_c}\mathcal{F}_{0,\pm}^{coll} \label{eq:coll0pm}
\end{equation}
and
\begin{equation}
\pm \frac{g_c \pi k_B^2T_\pm}{3 \hbar} \partial_xT_\pm \mp \frac{I F
}{2} =\mathcal{F}_{1,\pm}^{coll}\label{eq:coll1pm},
\end{equation}
respectively. The absence of thermal gradients in Eq.~\ref{eq:coll0pm} asserts the lack of thermoelectric power in systems with linear dispersion \cite{kuroda2008}. In Eqs.~\ref{eq:coll0pm} and \ref{eq:coll1pm}, $k_B$ is the Boltzmann constant and
\begin{equation}
\mathcal{F}_{m,\pm}^{coll} =\frac{g_c}{\pi \hbar v_F}
\int_{-\infty}^{\infty} \partial_t f_\pm  \Big\arrowvert_{coll}
(E-\mu)^m dE \label{eq:mom_coll_int}
\end{equation}
is the $m$-moment of the collision integral accounting for carrier scattering. We express the boundary conditions for $T_\pm(x)$ in terms of the transmission coefficient through the contacts $t$, such that:
\begin{equation}
T_\pm^2(\mp L/2)= t T_0^2 + (1-t) T_\mp^2(\mp L/2) \label{eq:electemp_bc}.
\end{equation}
where $T_0$ is the temperature at the contacts. The value of $t$ ranges between 0 and 1. The latter corresponds to perfect contacts that inject carriers from the leads without any reflection. After obtaining the $F$-profile (Eq.~\ref{eq:coll0pm}), the net drain source bias is computed as:
\begin{equation}
V_{ds} = I R_c - \int_{-L/2}^{L/2} F(x) dx
\end{equation}
where $R_c$ is the contact resistance which in the case of perfect contacts is $e^2/(\pi\hbar)$ \cite{imry}.

\subsection{Electron-phonon interaction}

The main contribution to the collision integrals (Eq.~\ref{eq:mom_coll_int}) in m-CNTs, other than e-e interaction that thermalizes the branch distributions, is scattering with lattice vibrations (phonons). The collision integrals read:
\begin{eqnarray}
\partial_t f_\eta  \Big\arrowvert_{\eta',\alpha-in} =R_{em,\alpha}^{\eta',\eta}
f_{\eta'}(\epsilon_{+\alpha}) \left[ 1\!
-\!f_\eta(E)\right]\nonumber\\
-R_{ab,\alpha}^{\eta,\eta'} f_{\eta}(E) \left[ 1\!-\!f_{\eta'}(\epsilon_{-\alpha})\right] \label{eq:coll_in}\\
\partial_t f_\eta  \Big\arrowvert_{\eta',\alpha-out} =R_{em,\alpha}^{\eta,\eta'} f_{\eta}(E) \left[ 1
-f_{\eta'}(\epsilon_{-\alpha})\right]-\nonumber\\
-R_{ab,\alpha}^{\eta',\eta} f_{\eta'}(\epsilon_{+\alpha}) \left[ 1\! -\!f_\eta(E)\right] \label{eq:coll_out}
\end{eqnarray}
where $\epsilon_{\pm\alpha} = E\pm\hbar\omega_\alpha$ and $R_{em,\alpha}^{\eta,\eta'}$ ($R_{ab,\alpha}^{\eta,\eta'}$) denotes the emission (absorption) scattering rate for the $\alpha$-phonon. Equation \ref{eq:coll_in} (Eq.~\ref{eq:coll_out}) is the collision integral for $\alpha$-phonons where the emission process scatters an electron from the carrier branch $\eta'$ ($\eta$) to the carrier branch $\eta$ ($\eta'$). These rates have different functional forms depending on the phonon branch involved in individual scattering events. The total contribution to carrier scattering with phonons is:
\begin{equation}
\partial_t f_\eta  \Big\arrowvert_{ph} = \sum_{\alpha,\eta'}\partial_t f_\eta  \Big\arrowvert_{\eta',\alpha-in}-\sum_{\alpha',\eta'}\partial_t f_\eta  \Big\arrowvert_{\eta',\alpha'-out}
\end{equation}

\begin{figure}
     \centering
     \includegraphics[width=3.25in]{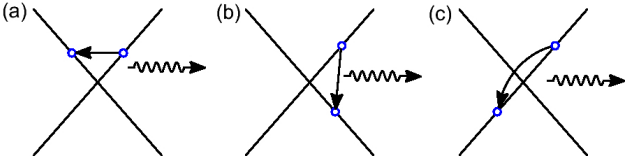}
     \caption{Phonon scattering mechanisms: (a) interbranch AP, (b) interbranch OP and (c) intrabranch OP.\label{fig:scatt_mechanisms} }
\end{figure}

For any given phonon wave-vector $q$ and energy $\hbar\omega_\alpha(q)$, the initial and final carrier wave-vector are determined by using momentum and energy conservation during collision. For instance with the Fermi energy located at the Dirac point ($k=0$ in Eq.~\ref{eq:bandst}), the conservation relations are written as:
\begin{eqnarray}
k &=& k' \pm q \label{eq:momcons}\\
E_\eta(k) &=& E_{\eta'}(k') \pm \hbar\omega_\alpha(q)\label{eq:encons}
\end{eqnarray}
where the + ($-$) sign denotes phonon emission (absorption) process from the $k$-state in the branch $\eta$ to the $k'$-state  in the branch $\eta'$. In our model we consider AP scattering, that dominates transport in the low-bias regime, and OP scattering that induces the electric transport nonlinearities in high-field. The different mechanisms are depicted in Fig.~\ref{fig:scatt_mechanisms}.

The AP populations are considered to be in thermal equilibrium with the lattice as their energies are smaller than thermal fluctuations at room temperature. In addition, we assume that carrier collisions with APs are elastic (i.e.~no energy is exchanged with the lattice but only between carrier populations) due to the small ratio between the sound and Fermi velocity ($v_s/v_F \lesssim 1/40$). Under these assumptions, the AP contribution to the collision integral vanishes when the initial and final states have the same Fermi velocity sign, and emission and absorption rates become equal. Therefore, the collision integral due to interbranch AP scattering is given by:
\begin{equation}
\partial_t f_\pm  \Big\arrowvert_{AP} = \sum_{\eta'} R_{AP} \left[f_\mp
(E)-f_\pm(E)\right],
\end{equation}
with $R_{AP} = (Z_A^2 k_B T)/(\rho v_s^2 \hbar ^2 v_F)$ by using the deformation potential approximation with $Z_A \sim$  5 eV \cite{park2004}. The first two moments of the collision integral for AP scattering read \cite{kuroda2008}:
\begin{eqnarray}
\mathcal{F}_{0,\pm}^{AP}= -\frac{g_c \left(\mu_+-\mu_-\right) }
{\pi \hbar v_F} \sum_{\eta'} R_{AP}^{\eta'} \label{eq:collint0ap}\\
\mathcal{F}_{1,\pm}^{AP}=\mp\frac{g_c \pi k_B^2}{6 \hbar v_F}
\left(T_+^2-T_-^2\right) \sum_{\eta'} R_{AP}^{\eta'}.
\label{eq:collint1ap}
\end{eqnarray}
Here we point out that any elastic backscattering, e.g.~scattering with impurities, would have given similar expressions with their specific rates, provided that they do not depend on carrier energies.

The OP emission and absorption scattering rates are written as:
\begin{equation}
R_{em,\alpha}^{\eta,\eta'} =
\frac{N_\alpha(q)+1}{\tau_{\alpha}^{ep}}\label{eq:em_scatt_rate}
\end{equation}
and
\begin{equation}
R_{ab,\alpha}^{\eta,\eta'} =
\frac{N_\alpha(q)}{\tau_{\alpha}^{ep}} \label{eq:abs_scatt_rate},
\end{equation}
respectively. In Eqs.~\ref{eq:em_scatt_rate} and \ref{eq:abs_scatt_rate}, $\tau_{\alpha}^{ep}$ and $N_\alpha(q)$ are the bare relaxation time and the occupation number for the $\alpha$-phonon, respectively. The experimental estimate for the former parameters is about 30 fs \cite{park2004,javey2004} and is an order of magnitude smaller than those values obtained by first principle calculations \cite{lazzeri2005}. We distinguish between interbranch ($\eta \neq \eta'$) and intrabranch ($\eta = \eta'$) OP scattering, as depicted in Fig.~\ref{fig:scatt_mechanisms}, and consider two interbranch (ZB and OP$_1$) and one intrabranch (OP$_2$) phonon modes interacting with carriers in m-CNTs \cite{auer2006}. Their energies are assumed to be constant and given by $\hbar\omega_{Z\!B} = $0.16eV (zone-boundary phonon) and $\hbar\omega_{O\!P_1} =\hbar\omega_{O\!P_2} = $0.20eV (interbranch  and intrabranch optical phonon), respectively.

If the OPs are in thermal equilibrium with the lattice, $N_\alpha(q)=N_{eq} = 1/[\exp(\hbar\omega_\alpha(q)/k_BT)-1]$ as given by the Bose-Einstein distribution, where $T=T(x)$ is the local lattice temperature. Otherwise, the occupation number $N_\alpha(q)$  depends on the local temperatures of carriers and lattice as well as the current level. At steady state, the occupation number of the $\alpha$-phonon is obtained by using the following rate equation:
\begin{eqnarray}
\partial_t N_\alpha(q) \Big\arrowvert_{\alpha}^{ep}+ \partial_t N_\alpha(q) \Big\arrowvert_{\alpha}^{pp}=0 \label{eq:phononBTE}
\end{eqnarray}
In Eq.~\ref{eq:phononBTE} we neglect phonon diffusion because of the small OP group velocity.
The first term in Eq.~\ref{eq:phononBTE} accounts for the generation/annihilation of OP due to e-p scattering and is given by:
\begin{eqnarray}
\partial_t N_\alpha(q) \Big\arrowvert_{\alpha}^{ep} = R_{em,\alpha}^{\eta,\eta'}f_{\eta}(\epsilon_{+\alpha}) \left[ 1\!
-\!f_{\eta'}(E)\right]\nonumber\\
-R_{ab,\alpha}^{\eta',\eta}f_{\eta'}(E) \left[ 1\!
-\!f_{\eta}(\epsilon_{-\alpha})\right] \label{eq:phonon_coll_int}
\end{eqnarray}
where $E$ is determined by using Eqs.~\ref{eq:momcons} and \ref{eq:encons}. The second term in Eq.~\ref{eq:phononBTE} describes the decay of OPs into APs and is expressed as:
\begin{equation}
\partial_t N_\alpha(q) \Big\arrowvert_\alpha^{pp} = -\frac{1}{\tau_{\alpha}^{pp}} \left[N_\alpha(q) - N_{eq}(q)\right] \label{eq:phdecay}
\end{equation}
where OP decay time $\tau_{\alpha}^{pp}$ has been estimated to be 1-7ps from Raman studies in m-CNTs \cite{oron2008, kang2008,bushmaker2007}. In addition, decay times have been found to vary as the inverse of $T$ \cite{pennington2008,kang2008}.

For intrabranch ($\eta = \eta'$) OP scattering, the phonon wave-vectors are $q = \pm \omega_{O\!P_2}/v_F$, independently of the initial or final state. The moments of the collision integral become:
\begin{eqnarray}
\mathcal{F}_{0,\pm}^{O\!P_2}=0 \\
\mathcal{F}_{1,\pm}^{O\!P_2}= \nonumber \frac{g_c^2 (\hbar\omega_{O\!P_2})^2}{\pi \hbar
v_F\tau_{O\!P_2}}\times\\
 \left[\frac{N_{O\!P_2}(q)+1}{\exp\left(\frac{\hbar\omega_{O\!P_2}}{k_BT_\pm}\right)-1}-
\frac{N_{O\!P_2}(q)}{\exp\left(-\frac{\hbar\omega_{O\!P_2}}{k_BT_\pm}\right)-1}\right].
\end{eqnarray}
For interbranch scattering with ZB and OP$_1$ modes, the phonon wave-vector $q$ depends on the initial and final electron states. Consequently, $N_\alpha(q)$ is obtained from Eq.~\ref{eq:phononBTE}, and depends on the carrier and lattice temperatures as well as the quasi-Fermi level separation (current). For $\tau_{\alpha}^{pp}\neq 0$ and $T_+\neq T_-$, the moments of the collision integrals (Eq.~\ref{eq:mom_coll_int}) for interbranch OP scattering do not have an analytical expression and are obtained numerically. In the Appendix \ref{app:aninteg}, we derive the expressions for the moments of the collision integral corresponding to interbranch scattering with OPs when $\tau_{\alpha}^{pp} = 0$ and $T_+ = T_-$.

\subsection{Electron-phonon heat exchange}

Combining the 1st moment equation for forward and backward carrier populations (Eq.~\ref{eq:coll1pm}), the heat production/dissipation satisfies:
\begin{equation}
\dot{q}_{el}+\dot{q}_{lat} = I F \equiv \dot{q} \label{eq:tot_heat_flow},
\end{equation}
where
\begin{equation}
\dot{q}_{el} \equiv \pm g_c G_{th}^+ \partial_xT_+ + g_c G_{th}^-\partial_xT_- \label{eq:el_heat_flow}
\end{equation}
is the amount of heat carried by the electrons and
\begin{equation}
\dot{q}_{lat} \equiv \frac{1}{2\pi}\sum_\alpha \int_{-\infty}^\infty \hbar\omega_\alpha(q)\, \partial_t N_\alpha(q)\Big\arrowvert_{\alpha}^{ep} dq
\end{equation}
is the amount of energy removed from the carriers for the OP populations. In addition, since heat transport is neglected in Eq.~\ref{eq:phdecay}, $\dot{q}_{lat}$ denotes the rate (per unit length) of energy transferred to the lattice (APs) by OP decays. This value determines the local lattice heating, and depends on the current and the carrier and lattice temperatures. We assume that lattice (AP) heat transport is diffusive and follows the Fourier's law for which the heat equation reads:
\begin{equation}
\dot{q}_{ap}+ \dot{q}_{sub} = \dot{q}_{lat}\label{eq:lat_heat_flow}.
\end{equation}
The first term in the left hand side of Eq.~\ref{eq:lat_heat_flow} accounts for the heat carried by the AP population,
\begin{equation}
\dot{q}_{ap} \equiv -A \partial_x\left[\kappa(T) \partial_x T\right] \label{eq:ap_heat_flow},
\end{equation}
The factor $A = \pi d t$ is the cross sectional area, where $d$ and $t\approx 0.34$nm are the diameter and the thickness of the CNT, respectively. The lattice thermal conductivity $\kappa(T)$ is given by $\kappa(T) = \kappa_0 T_0 /T$ due to the Umklapp phonon-phonon scattering \cite{osman2001}. The room temperature thermal conductivity has been estimated to be between 15 and 60 W/(cm K) for SWCNTs \cite{hone1999}. The second term in Eq.~\ref{eq:lat_heat_flow} determines the heat removal through the substrate, which is modeled as a contact resistance \cite{durkan1999}:
\begin{equation}
\dot{q}_{sub} \equiv g_0 \left(T-T_0\right) \label{eq:subs_heat_flow}.
\end{equation}
The parameter $g_0$ denotes the thermal coupling between the CNT and the substrate which has been experimentally estimated between 0.05 and 0.20 W/(Km) for SiO$_2$\cite{maune2006,pop2007}. The lattice temperature boundary conditions between CNT and leads are set by the presence of a contact resistance:
\begin{equation}
\kappa \partial_xT\big|_{x=\pm L/2} = \frac{T(\pm L/2) - T_0}{\mathcal{R}_{th}}\label{eq:lattemp_bc}
\end{equation}
with $\mathcal{R}_{th}\sim 10^7$K/W for m-CNTs\cite{pop2008}.

\subsection{Heat flow diagram}

\begin{figure}[tb]
    \centering
    \includegraphics[width=3.25in]{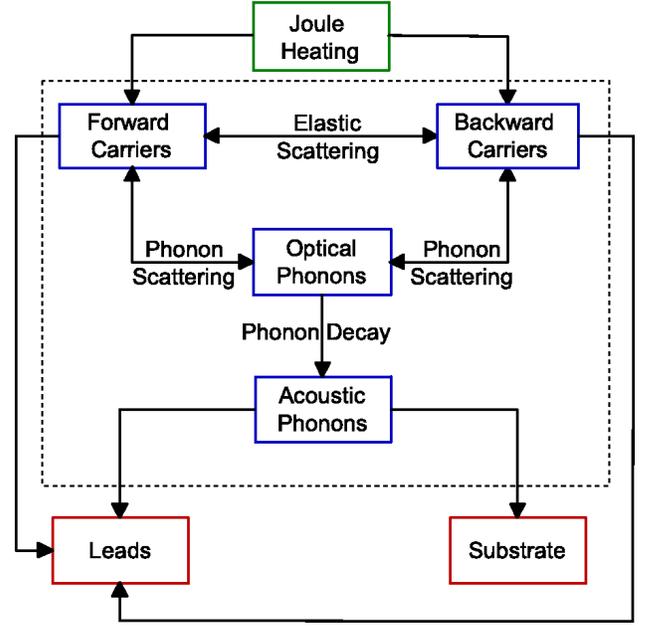}
    \caption{Flowchart of the heat transfer. The power given to the system by the external field is transferred to the forward and backward carrier populations. Simultaneously, these populations exchange heat with OP populations by emission/absorption processes. In addition OPs decay into APs, transferring heat to the lattice. The heat transferred from the field is removed through the leads or the substrate. \label{fig:heatflow}}
\end{figure}

The coupled nonlinear integro-differential equations for the carrier and lattice temperatures requires a self-consistent solution of Eqs.~\ref{eq:coll1pm} and \ref{eq:lat_heat_flow}. The interplay between carrier and phonon heat transfer (Eq.~\ref{eq:tot_heat_flow}) is depicted in Fig.~\ref{fig:heatflow}, where both forward and backward carrier populations gain energy from the electric field. In the high-field regime, OP population builds up (hot phonon effects) due to the enhanced scattering by phonon emission, but because of their small group velocity, heat removal by OPs is not significant. Consequently, either they transfer their energy back to the carrier population in reabsorption processes, or they decay into APs, thereby heating the lattice. Energy can be removed from the system through the leads (by both carriers and APs) or by the substrate. A key factor that characterizes heat transfer is the ratio $\tau_{\alpha}^{ep}/\tau_{\alpha}^{pp}$ which determines the fraction of heat transferred from the lattice \cite{steiner2009, vandecasteele2009}. On the one hand if $\tau_{\alpha}^{ep}/\tau_{\alpha}^{pp} \gg 1$, most of the energy gained by the OP population is transferred to the APs by instantaneous phonon decay and carried by the lattice. On the other hand, if $\tau_{\alpha}^{ep}/\tau_{\alpha}^{pp} \ll 1$, the energy of an OP emitted by carrier scattering is more likely to return to the carrier population by reabsorption rather than decay into APs. Hence, most of the energy remains within the carrier and OP populations. However, since OPs cannot carry significant amount of heat (because of their small group velocity), heat is removed by the electron populations through the leads. In this case, the presence of a substrate would make no difference -- considering that $\tau_{\alpha}^{ep}$ or $\tau_{\alpha}^{pp}$ are not affected by the presence of the substrate -- because only a small fraction of heat is transferred to the lattice.

\section{Results}

\subsection{Low-field regime}

\begin{figure}[htpb]
    \centering
    \includegraphics[width=3.25in]{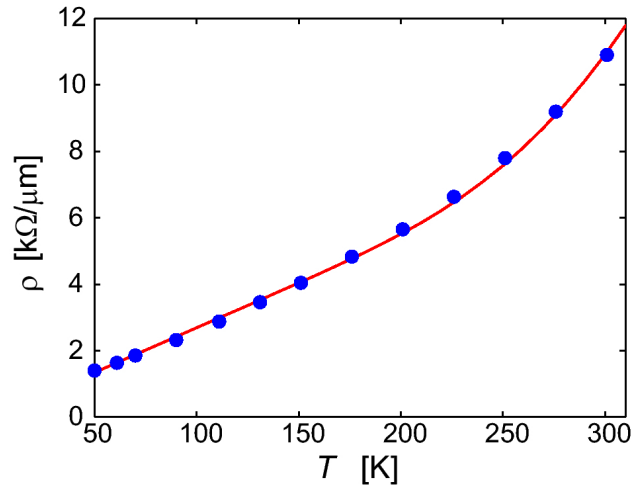}
    \caption{Temperature dependence of the low-field resistivity: theory (solid lines) and experiment (symbols) from Purewal et al.~\cite{purewal2007}.}\label{fig:low_field_resist}
\end{figure}

Heating of carrier or lattice population is negligible ($|T_\pm(x)-T_0|,|T(x)-T_0|\ll T_0$) in the low-field regime. In this case, the collision integrals for OP scattering have an analytic form (see Appendix). Combining Eqs.~\ref{eq:coll0pm} with \ref{eq:collint0ap} and \ref{eq:collint0eqm}, we obtain the Matthiessen rule for the low-field resistivity:
\begin{equation}
\rho(T) = \frac{1}{2G_0}\left(\frac{1}{\lambda_{AP}^{e\!f\!f}}+  \frac{1}{\lambda_{OP}^{e\!f\!f}}\right)\label{eq:resistivity_lf}
\end{equation}
where $\lambda_{A\!P}^{e\!f\!f}$ and $\lambda_{O\!P}^{e\!f\!f}$ are the effective AP and OP mean free path given by:
\begin{equation}
\lambda_{AP}^{e\!f\!f} =  \frac{T_0\lambda_{AP,0}}{T}
\end{equation}
and
\begin{equation}
\lambda_{op}^{e\!f\!f}=\left\{\sum_\alpha \frac{2}{v_F\tau_{\alpha}^{ep}} \frac{\hbar \omega_{\alpha}/(2 k_B T)}{\sinh\left[\hbar \omega_{\alpha}/(2 k_B T)\right]^2}\right\}^{-1}\label{eq:mfp_lf}
\end{equation}
This analytical expression is in very good agreement with the temperature dependence of the low-field resistivity in m-CNT obtained experimentally~\cite{purewal2007} for the parameters $\lambda_{A\!P,0}$ = 800nm, $\tau_{Z\!B}$ = 60fs, $\tau_{O\!P_1}$ = 100fs $\hbar\omega_{Z\!B}$= 0.16eV, $\hbar\omega_{O\!P_1}$= 0.20eV, and $T_0 = 300K$ as shown in Fig.~\ref{fig:low_field_resist}.

\begin{figure}[tb]
    \centering
    \includegraphics[width=3.25in]{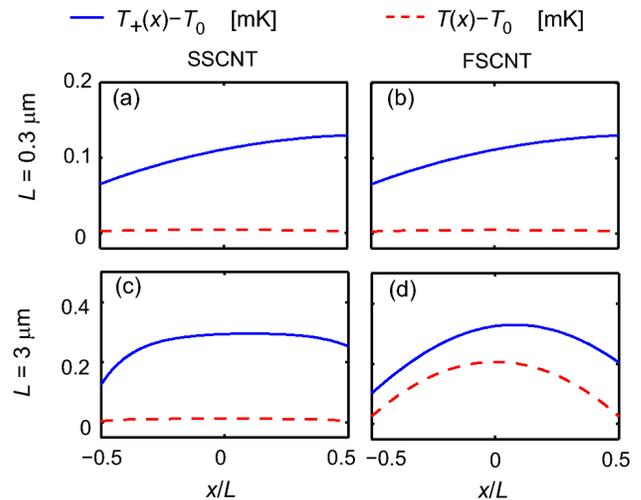}
    \caption{Forward carrier (solid line) and lattice (dashed line) temperature difference to the contact temperature $T_0$ corresponding for $I=0.01$ $\mu$A (low-field regime): (a) 0.3 $\mu$m-long SSCNT,  (b) 0.3 $\mu$m-long FSCNT, (c) 3 $\mu$m-long SSCNT, and (d) 3 $\mu$m-long FSCNT.\label{fig:iv_comparison_lf}}
\end{figure}

In Fig.~\ref{fig:iv_comparison_lf} we plot the forward carrier (solid line) and lattice (dashed line) temperature profiles corresponding to the low-field regime ($I$ = 0.01$\mu$A) for 0.3 and 3$\mu$m-long m-CNTs. These profiles are determined for both SS and FSCNT configuration, for which the thermal coupling to the substrate are $g_0 \approx1.5\times10^{-3}$W/(cm K) and 0, respectively. In the case of symmetric boundary conditions for carrier temperatures, i.e.~$T_+(-L/2) = T_-(L/2)$, and lattice temperature, i.e.~$T(-L/2) = T(L/2)$, the carrier temperature profiles satisfy $T_+(x) = T_-(-x)$ due to the electron-hole symmetry, thereby reducing the number of differential equations. In this regime the electric field along the channel remains constant. For short CNTs the profiles are similar regardless of the presence/absence of substrate with the heating in the carrier population being more significant than in the lattice. The forward carrier temperature increases along the tube, and the maximum is reached at the drain as a consequence of the long mean free path (compared to the CNT length) \cite{kuroda2009}. In the case of long m-CNTs, despite of substantial inelastic scattering that heats the AP population, the carrier temperature increase is larger than that of the lattice. In particular, for FSCNT lattice heating is comparable to that of the carriers, while in the SSCNT the lattice temperature is flattened by the efficient heat removal through the substrate, and remains closer to the substrate temperature. The maximum lattice temperature is located exactly at the m-CNT mid-length, while the forward carrier temperature peak is shifted towards the right lead.

\subsection{High-field regime}

\begin{figure}[htpb]
    \centering
    \includegraphics[width=2.7in]{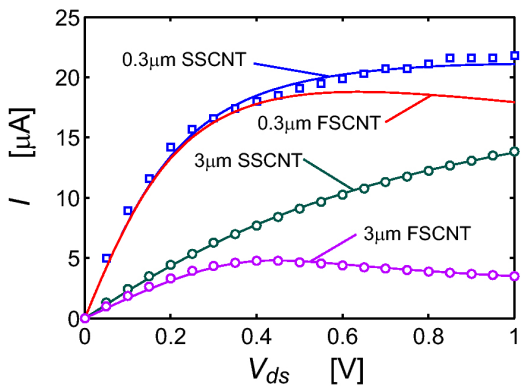}
    \caption{$IV$ characteristics of 0.3 and 3$\mu$m-long CNTs in both SS and FS configurations. Symbols correspond to experimental data from Javey et al.~\cite{javey2004} ($\square$) and Pop et al.~\citep{pop2005} ($\bigcirc$).\label{fig:iv_comparison}}
\end{figure}

In Fig.~\ref{fig:iv_comparison} we display the electrical characteristics (solid lines) corresponding to the SS and FS configurations of 0.3 and 3$\mu$m-long CNTs that are in excellent agreement with experiments~\cite{javey2004,pop2005}. The parameters used in the calculations are: $\kappa$(300K) = 20 W/(cm K), $g_0=1.5\times10^{-4} \mbox{W/(cm K)}$ and $Z_A \approx 5$ eV. The relaxation times $\tau_{Z\!B} \approx 85$fs, $\tau_{O\!P_1}=\tau_{O\!P_2} \approx 200\mbox{fs}$ and $\tau_{\alpha}^{pp} = 1\mbox{ps}$ are within the range estimated by experiments \cite{javey2004,pop2005} and theory \cite{lazzeri2005}. Nonlinear electrical characteristics are obtained for all CNTs. In the high-bias regime, the CNTs in FS configuration exhibit negative differential resistance (NDR) induced by the inefficient heat removal that increases scattering. However, while in 0.3 $\mu$m-long FS and SS CNTs the current levels approach the 20 $\mu$A limit attributed to the onset of OP emission \cite{yao2000}, in the 3 $\mu$m-long FSCNT the current level is substantially reduced (but still exhibiting NDR features). By contrast in the 3 $\mu$m-long SSCNT, the current increases monotonically in the bias range shown, and no saturation is observed due to the fact that the electric field is about an order of magnitude smaller than in the 0.3 $\mu$m-long SSCNT for the same voltages.

\begin{figure}
    \centering
    \includegraphics[width=3.25in]{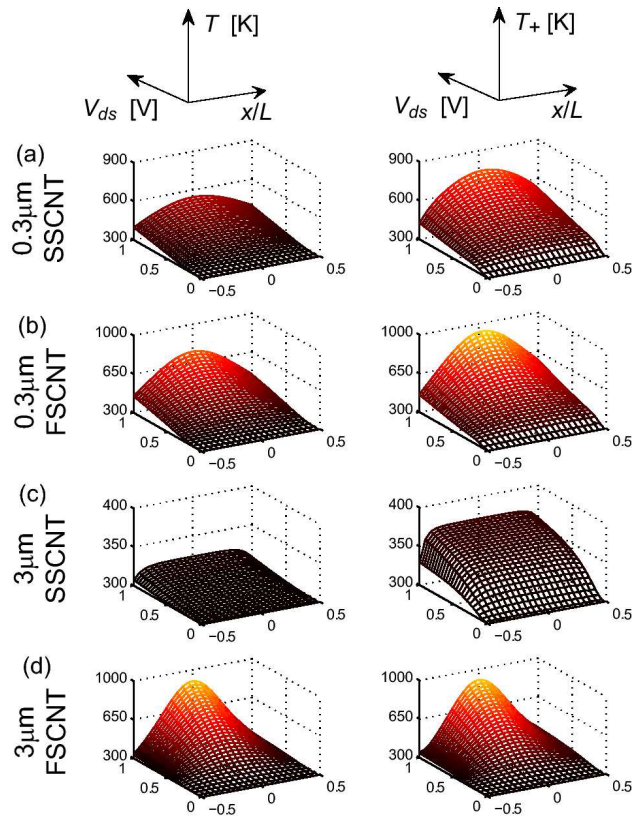}
    \caption{Lattice (left) and carrier (right) temperature profile as a function of the bias voltage $V_{ds}$ for the $IV$ characteristics shown in Fig.~\ref{fig:iv_comparison}. From top to bottom: 0.3$\mu$m-long SSCNT, 0.3$\mu$m-long FSCNT, 3$\mu$m-long SSCNT, and 3$\mu$m-long FSCNT. \label{fig:temp_prof}}
\end{figure}

The corresponding lattice and forward carrier temperature profiles are shown in Fig.~\ref{fig:temp_prof}. For short (0.3$\mu$m-long) m-CNTs, the temperature profiles in both FS and SS configurations have a parabolic shape, as shown in Fig.~\ref{fig:temp_prof}(a) and \ref{fig:temp_prof}(b). Unlike the low-field regime (Fig.~\ref{fig:iv_comparison_lf}), the temperature maxima for forward carriers in the high-field regime occur at the center of the m-CNT in both cases, in agreement with breakdown experiments in m-CNTs \cite{collins2001,chiu2005}. We note that for the same channel length, the values of the carrier and lattice temperature maxima are lower in the SSCNT than in the FSCNT as heat is removed by the substrate. In SSCNTs, the reduced temperature lowers scattering rates, and consequently higher current levels are achieved compared to their FS counterparts. By contrast, in 3$\mu$m-long SSCNTs the efficient heat removal through the substrate flattens the temperature profiles as depicted in Fig.~\ref{fig:temp_prof}(c), as observed recently \cite{shi2009}. Because of the electron-hole symmetry ($T_+(x)=T_-(-x)$) forward and backward carrier temperatures are equal to one another in the channel, but are different in the regions close to the contacts. However due to the effective heat transfer to the substrate, carrier populations are not locally in thermal equilibrium with the lattice. For the 3 $\mu$m-long FSCNT (Fig.~\ref{fig:temp_prof}.d), carrier and lattice temperature profiles have a parabolic shape and markedly higher values due to limited heat removal, which enhances carrier scattering and phonon decays. The current level reduction as well as the near thermal equilibrium between carrier and phonon populations indicates that transport approaches the diffusive regime limit \cite{nagaev1995}.

\begin{figure*}[tbp]
    \includegraphics[width=6in]{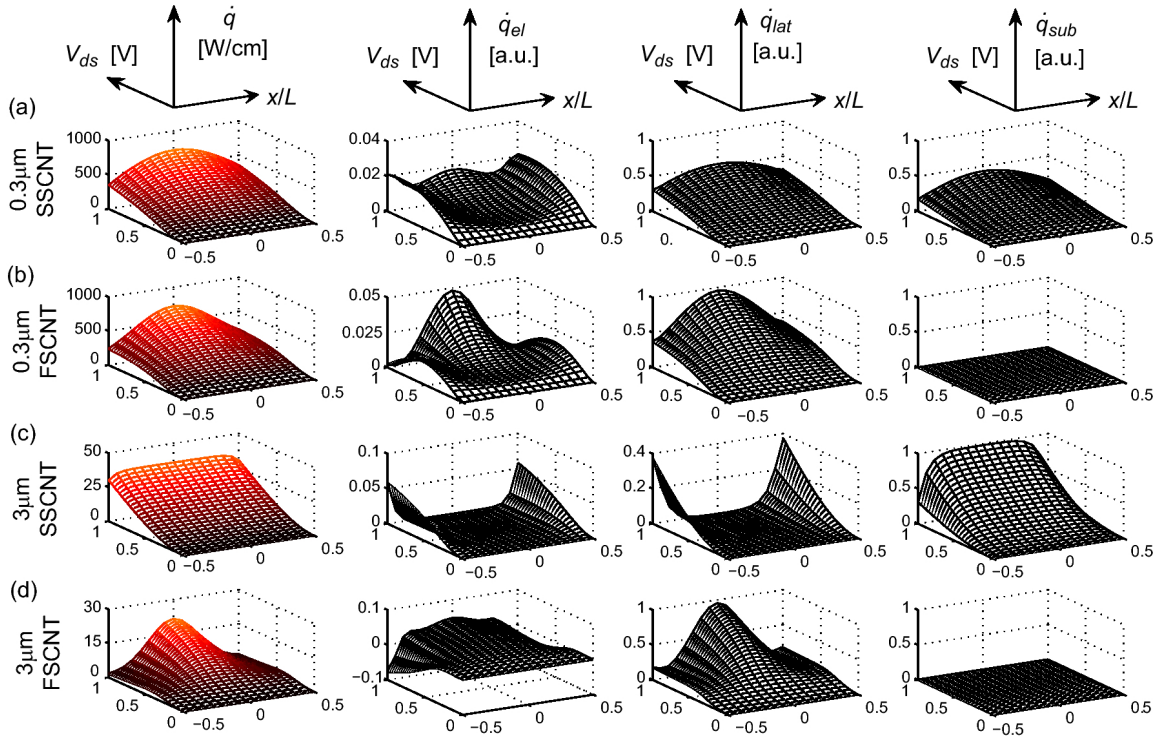}
    \caption{2D heat flow profiles as a function of $V_{ds}$ and distance for 0.3 and 3$\mu$m-long CNTs, for the $IV$ characteristics and temperature profiles depicted in Figs.~\ref{fig:iv_comparison} and \ref{fig:temp_prof}, respectively. From top to bottom: 0.3$\mu$m-long SSCNT, 0.3$\mu$m-long FSCNT, 3$\mu$m-long SSCNT, and 3$\mu$m-long FSCNT. First column corresponds to the Joule's heat production ($\dot q_{tot}$). Second, third and fourth column display the fraction of heat carried by the electrons ($\dot{q}_{el}$), acoustic phonons ($\dot{q}_{ap}$), and substrate ($\dot{q}_{sub}$), respectively. The profiles are normalized to the maximum power density across the channel at $V_{ds}$=1V. \label{fig:htflw}}
\end{figure*}

\subsection{Heat flow} \label{sn:htflw}

In Fig.~\ref{fig:htflw} we simultaneously display the local heat production given by Joule's law (Eq.~\ref{eq:tot_heat_flow}) and the fraction of the local heat carried by electrons ($\dot{q}_{el}$ in Eq.~\ref{eq:el_heat_flow}), APs ($\dot{q}_{ap}$ in Eq.~\ref{eq:lat_heat_flow}), and substrate ($\dot{q}_{sub}$ in Eq.~\ref{eq:subs_heat_flow}) for the $IV$ characteristics shown in Fig.~\ref{fig:iv_comparison}. For illustrative purposes, the profiles have been normalized to the maximum power per unit length across the channel for $V_{ds}=1$V. Since the current remains constant along the channel the $\dot{q}$-profile is proportional to the electric field profile and presents a peak at the middle of the m-CNT. Despite significant carrier heating (left column of Fig.~\ref{fig:temp_prof}), the fraction of heat carried by electrons (second column in Fig.~\ref{fig:htflw}) is small ($<10\%$) compared to the total heat produced locally for all cases shown. Indeed, most of the heat is removed by APs (lattice) and/or the substrate (in the SS configuration), as depicted in the third and fourth columns of Fig.~\ref{fig:htflw}. For 0.3 $\mu$m-long CNTs, the heat production profiles in the SS and FS configurations look alike despite the different heat removal mechanisms, as shown in Fig.~\ref{fig:htflw}.a and Fig.~\ref{fig:htflw}.b. In contrast for 3 $\mu$m-long m-CNTs, the field is uniform in the SS configuration (deviations are less 10\% for biases shown), whereas it is and highly inhomogeneous with its peak value at the CNT midlength in the FS configuration. We notice that, in sufficiently long SSCNTs where the substrate flattens the temperature profile, the local field value does not depend on the boundary conditions. In this case, an approximate solution can be obtained by solving for the values of carrier and lattice temperatures such that their corresponding temperature gradients vanish in the coupled differential equations (Eqs.~\ref{eq:coll0pm}, \ref{eq:coll1pm} and \ref{eq:lat_heat_flow}). This self-consistent analysis of the thermal flow shows that the carrier contribution to heat transport in m-CNTs with $L\gtrsim 300$nm can be neglected and, therefore, validates the usual approximation \cite{kuroda2005,pop2005} $\dot{q}_{lat} \approx IF$.

We point out that recent works \cite{perebeinos2009,rotkin2009} have suggested the existence of surface polariton scattering in the SS-configuration. This scattering mechanism would indeed modify the heat transport scheme (Fig.~\ref{fig:heatflow}) by adding a new carrier-substrate channel to heat removal (without the mediation of APs), which could be accounted for by including an additional term on the left hand side of Eq.~\ref{eq:tot_heat_flow}. However, the strength of this mechanism, which decays as a power law with the CNT-substrate distance, and possibly varies with gate bias, still requires confirmation.

\subsection{Hot Phonons}

\begin{figure}[htpb]
    \centering
    \includegraphics[width=3.25in]{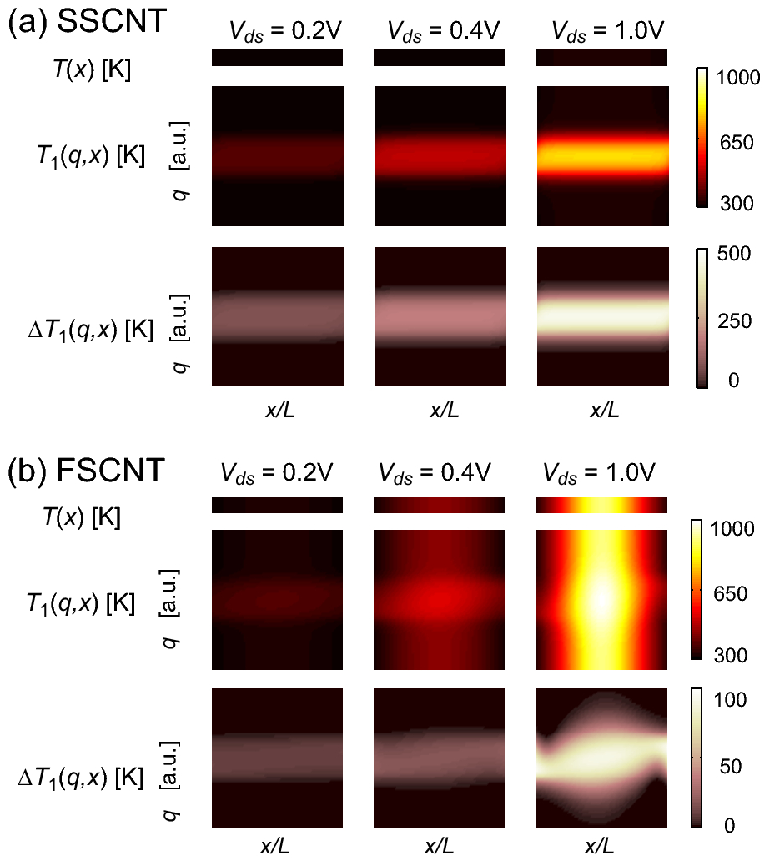}
    \caption{2D contour plot of the non-equilibrium ZB phonon equivalent temperature as a function of the phonon wave vector $q$ and distance for a 3$\mu$m-long CNT in: (a) SS and (b) FS configuration under various external biases: 0.20V (left), 0.40V (center), and 1.00V (right). For each bias the lattice temperature $T(x)$ profile along the channel, the effective temperature of the first phonon mode $T_{Z\!B}(x,q)$ and the temperature difference $\Delta T_{Z\!B}(x,q)$ between each mode and the lattice temperature are shown in the top, middle and bottom color plots, respectively. Note that $T_{Z\!B}(x,q)$ are plotted using the same color scale for both lengths but $\Delta T_{Z\!B}(x,q)$ has been represented with different ones. \label{fig:hot_phonons} }
\end{figure}

By solving Eq.~\ref{eq:phdecay} self-consistently with the carrier and lattice temperature profiles for a particular voltage we compute the OP occupation number $N_\alpha(q)$ and determine the equivalent of the temperature of the $\alpha$-phonon mode $T_\alpha(q)$ as:
\begin{equation}
T_{\alpha}(q) = \frac{\hbar\omega_{\alpha}} {k_B \log \left[1+\frac{1}{N_{\alpha}(q)}\right]}\label{eq:optemp}.
\end{equation}
In Fig.~\ref{fig:hot_phonons}, we plot the lattice temperature profile $T(x)$ (narrow surface plot), the temperature $T_{Z\!B}(q,x)$ corresponding to the ZB phonon mode, and the difference between the phonon mode temperature and the local temperature profile, i.e.~$\Delta T_{Z\!B}(q,x) = T_{Z\!B}(q,x)-T(x)$ for $V_{ds}$ values 0.2, 0.4, and 1V in the 3 $\mu$m-long CNTs for which we have previously computed the electrical characteristics and corresponding temperature profiles (Figs.~\ref{fig:iv_comparison} and \ref{fig:temp_prof}). On the one hand we observe that the phonon temperature profile in SSCNT is uniform along the $x$-direction (except for the regions close to the leads) because of the constant carrier and lattice temperature profiles as displayed in Fig.~\ref{fig:hot_phonons}(a). 
For $V_{ds}=1\mbox{V}$, the high current levels (i.e.~large quasi-Fermi level separation) favor scattering by OP emission \cite{kuroda2006} and a non-equilibrium OP population builds up (due to non-zero $\tau_{\alpha}^{pp}$). On the other hand, the OP equivalent temperature profiles in the 3 $\mu$m-long FSCNT are not uniform reaching peak values close to 1000 K for $V_{ds}=1\mbox{V}$. Nonetheless, in the FS configuration the deviations $\Delta T_{Z\!B}$ are smaller than in the SS one, as shown (in different color scales) at the bottom of Fig.~\ref{fig:hot_phonons}(a) and Fig.~\ref{fig:hot_phonons}(b), respectively. The high lattice temperature achieved in FSCNTs increases both emission and absorption scattering rates, thereby damping (non-equilibrium) hot phonon effects as carrier and phonon populations reach local thermal equilibrium ($T_\pm(x) = T(x) = T_{OP}(x)$). Therefore, in the absence of hot phonon effects, our model shows that the cause for the onset of NDR at high bias is, as anticipated by Conwell \cite{conwell2008}, the existence of the electric field inhomogeneities along the heated CNT \cite{kuroda2006}.

\subsection{Electrical power vs length}

\begin{figure}
    \centering
    \includegraphics[width=3.25in]{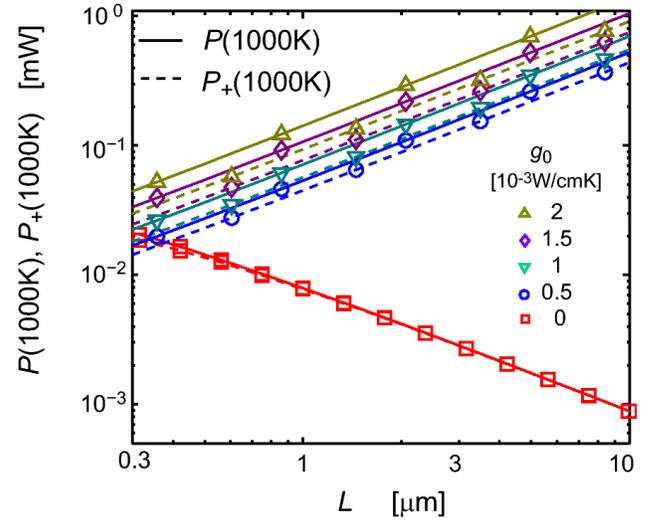}
    \caption{Electrical powers, $P(1000\mbox{ K})$ and $P_+(1000\mbox{ K})$, dissipated in the CNT when length when the maximum lattice (solid lines) and carrier (dashed lines) temperatures reach 1000K, respectively. The symbols indicate the different values of the thermal coupling to the substrate. \label{fig:brkdwn} }
\end{figure}

Electrical breakdown by oxidation in CNTs has been estimated to occur at around 1000K \cite{radosavljevi2001,cataldo2002}. On Fig.~\ref{fig:brkdwn} we display the electrical powers $P(1000\mbox{ K})$ (solid line) and $P_+(1000\mbox{ K})$ (dashed line) as a function of the CNT length for various values of the thermal coupling to the substrate $g_0$. The parameters $P(1000\mbox{ K})$ and $P_+(1000\mbox{ K})$ correspond to the electrical power ($P = IV$) dissipated along the CNT when the maximum lattice temperature and maximum carrier temperature reach 1000K, respectively. For the FSCNT ($g_0 = 0$), both electrical powers are determined from the self-consistent solution of the BTE and heat equation. For SSCNTs ($g_0>0$), we compute the power as $P = I (I R_q + F L)$ by assuming that the electric field is homogeneous along the conductor (vanishing thermal gradients) and solving for the values of electric field and currents as described in Section \ref{sn:htflw}. In the former case, i.e.~$g_0=0$ ($\square$), $P(1000\mbox{ K})$ and $P_+(1000\mbox{ K})$ are approximately equal (both curves are indistinguishable when $L\gtrsim 1 \mu$m), and scale inversely proportional to the length. In contrast the dissipated powers in SSCNTs increase linearly with length for sufficiently long m-CNTs, when the voltage drop at the contacts is smaller than that along the channel. Similar trends have been reported experimentally in the electrical breakdown of multi-walled CNTs in both FS and SS configuration \cite{chiu2005}. We point out that at a fixed length $P_+(1000\mbox{ K})$ and $P(1000\mbox{ K})$ increase with the substrate coupling. Additionally, the difference $\Delta P = P(1000\mbox{ K})-P_+(1000\mbox{ K})$ widens by the enhancement of $g_0$ indicating more prominent hot electron effects as carrier populations reach the 1000 K temperature at lower power values than the lattice.

\section{Conclusions}

A model for electro-thermal transport in 1D metallic systems in the high temperature (incoherent) regime has been presented. We use the BTE formalism to describe the interdependent electron-phonon dynamics in high-field. By accounting for the emergence of carrier and phonon populations out of equilibrium we quantify the heat production and dissipation by each of these populations depending on the experimental configuration. The phonon decay rate is the bottleneck that regulates heat exchange between carriers and lattice. Our model shows remarkable agreement with high-field transport experiments and offers a qualitative interpretation of breakdown experiments in multi-walled-CNTs. We attribute the emergence of negative differential resistance observed in suspended CNTs to the inhomogeneous electric field and self-heating in the high field regime.

One of the authors (MK) acknowledges the support of the Department of Physics and the Department of Electrical and Computer Engineering at the University of Illinois at Urbana-Champaign.

\appendix*
\section{Analytical expressions for the collision integrals} \label{app:aninteg}

For carriers in different branches in thermal equilibrium (i.e. $T_+ = T_-\equiv T_{el}$), and when hot phonon effects are neglected (i.e. $\tau_{ep} \gg \tau_{\alpha}^{pp}$ or low current level), analytical expressions for the zeroth and first moments of the collision integral (Eq.~\ref{eq:mom_coll_int}) and the OP generation by phonon emission/absorption (Eq.~\ref{eq:phonon_coll_int}) can be obtained:
\begin{widetext}
\begin{eqnarray}
\mathcal{F}_{0,\pm}^{\alpha}=\pm \frac{g_c^2 k_BT_{el}}{\hbar \pi v_F
\tau_{op}} \left\{ (N_\alpha(q)+1)
\left[\mathcal{G}(\Omega_+)-\mathcal{G}(\Omega_-)\right] +
N_\alpha(q)
\left[\mathcal{G}(-\Omega_-)-\mathcal{G}(-\Omega_+)\right]\right\}\label{eq:collint0eqm}
\\
\mathcal{F}_{1,\pm}^{\alpha}=-\frac{g_c^2
\hbar\omega_{\alpha}k_BT_{el}}{2\hbar \pi v_F \tau_{op}} \left\{
(N_\alpha(q)+1)
\left[\mathcal{G}(\Omega_+)+\mathcal{G}(\Omega_-)\right] -
N_\alpha(q)
\left[\mathcal{G}(-\Omega_-)+\mathcal{G}(-\Omega_+)\right]\right\} \label{eq:collint1eqm}\\
\partial_t N_\alpha \Big\arrowvert_{pp} = \frac{g_c^2 \hbar\omega_{\alpha}k_BT_{el}}{2\hbar \pi v_F \tau_{op}}
\left\{ (N_\alpha(q)+1)
\left[\mathcal{G}(\Omega_+)+\mathcal{G}(\Omega_-)\right]-N_\alpha(q)
\left[\mathcal{G}(-\Omega_-)+\mathcal{G}(-\Omega_+)\right]\right\}
\label{eq:hot_phononseqm}
\end{eqnarray}
\end{widetext}
where we define $\mathcal{G}(x) = \frac{x}{\exp(x)-1}$ and $\Omega_\pm = \frac{\hbar\omega_{op}\pm\Delta\mu}{k_BT_{el}}$. The first two expressions are valid in the low-bias regime, where heating in the carrier population and lattice can be neglected. In addition, the omission of hot phonon effects is valid approximation for the long FSCNTs because the minor thermal imbalance between OP and AP (lattice) populations.

\bibliography{report}

\end{document}